\documentclass[aps,prl,twocolumn,showpacs,preprintnumbers,amsmath,amssymb,superscriptaddress]{revtex4-2}
\usepackage{epsfig,amsopn}
\usepackage{graphicx}
\usepackage{epstopdf}
\usepackage{sidecap}
\usepackage{color}
\usepackage{amsmath,amssymb}
\usepackage{amsthm}
\usepackage{amsfonts}
\usepackage{textcomp}
\usepackage{enumerate}
\usepackage{soul}
\usepackage{float}
\usepackage{scalerel}
\usepackage{braket}
\usepackage{tikz}
\usepackage{natbib}
\usepackage[titletoc]{appendix}
\usepackage[ colorlinks = true,
linkcolor = red,
urlcolor  = red,
citecolor = blue,
anchorcolor = red]{hyperref}

\newcommand{\beq}{\begin{equation}}
  \newcommand{\eeq}{\end{equation}}
\newcommand{\bqa}{\begin{eqnarray}}
  \newcommand{\eqa}{\end{eqnarray}}

\begin{document}
	\title{ Andreev versus Tunneling Spectroscopy of Unconventional Flat Band Superconductors}
	  \author{Sayak Biswas}
        \affiliation{Department of Physics, The Ohio State University, Columbus, OH 43201, USA}
          \author{Saurav Suman}
          \affiliation{Department of Theoretical Physics, Tata Institute of Fundamental Research,
          	Homi Bhabha Road, Mumbai 400005, India }
		\author{Mohit Randeria}
		  \affiliation{Department of Physics, The Ohio State University, Columbus, OH 43201, USA}
		  \author{Rajdeep Sensarma}
	\affiliation{Department of Theoretical Physics, Tata Institute of Fundamental Research,
		Homi Bhabha Road, Mumbai 400005, India }
	%
	%
	%
	\begin{abstract}
STM experiments in the tunneling and Andreev regimes on graphene-based moiré superconductors (SC) show two distinct energy scales whose origin is mysterious.
We express the conductance of a normal-SC interface in terms of Green's functions, which allows us to sharpen the issues in two ways. 
First, we show that the two distinct energy scales cannot be understood in terms of a pseudogap in tunneling and a superconducting gap in the Andreev spectra. 
Second, the large Fermi velocity $v_F$ mismatch between the STM tip and the flat band SC renormalizes a transparent interface towards the tunneling regime, 
and the ballistic Andreev regime cannot be realized in moire SCs. We also discuss self energy corrections to $v_F$ that determines the conductance.
Finally, we offer a resolution to these problems by modeling the Andreev experiment as a circular metallic disc embedded in an unconventional 
SC. We show that with strong $v_F$ mismatch the low bias conductance is dominated by
Andreev bound states (ABS) induced by the tip at the interface with the unconventional SC. 
The ABS give rise to the low energy scale seen in Andreev experiments, smaller than the SC gap in tunneling spectroscopy.
	\end{abstract}
	\maketitle
 
The discovery of superconductivity (SC) in magic-angle twisted bilayer graphene 
(TBG)~\cite{cao2018,yankowitz2019,lu2019, arora2020,saito2020,andrei2020} and 
related moire materials \cite{hao2021} raises many challenging questions 
These include: What is the symmetry of the SC order parameter? 
What is the mechanism of SC in topological flat bands? What role, if any, do nearby correlated insulating states play? 
Scanning tunneling microscopy (STM) spectra has given insight into these questions, indicating nodes in the SC gap
in both TBG \cite{oh2021} and twisted trilayer graphene (TTG) \cite{kim2022}.
 This is also consistent with power-law temperature dependence of the superfluid stiffness in TBG~\cite{banerjee2025} and TTG~\cite{tanaka2025}. 

There are, however, puzzling differences between the STM results~\cite{oh2021,kim2022} in the tunneling regime with a large resistance between the tip 
and the sample, and the low resistance Andreev regime where the tip is actually touching the sample. 
The two experiments reveal rather different characteristic energy scales, differing by as much as a factor of $\sim 4$.
The larger energy scale seen in tunneling has been tentatively identified with a pseudogap scale, while the one seen in the Andreev regime,
has been associated with a coherent SC gap. This idea traces its roots to early work on cuprates~\cite{deutscher1999, deutscher2005}.

\begin{figure}
    \centering
    \includegraphics[width=\linewidth]{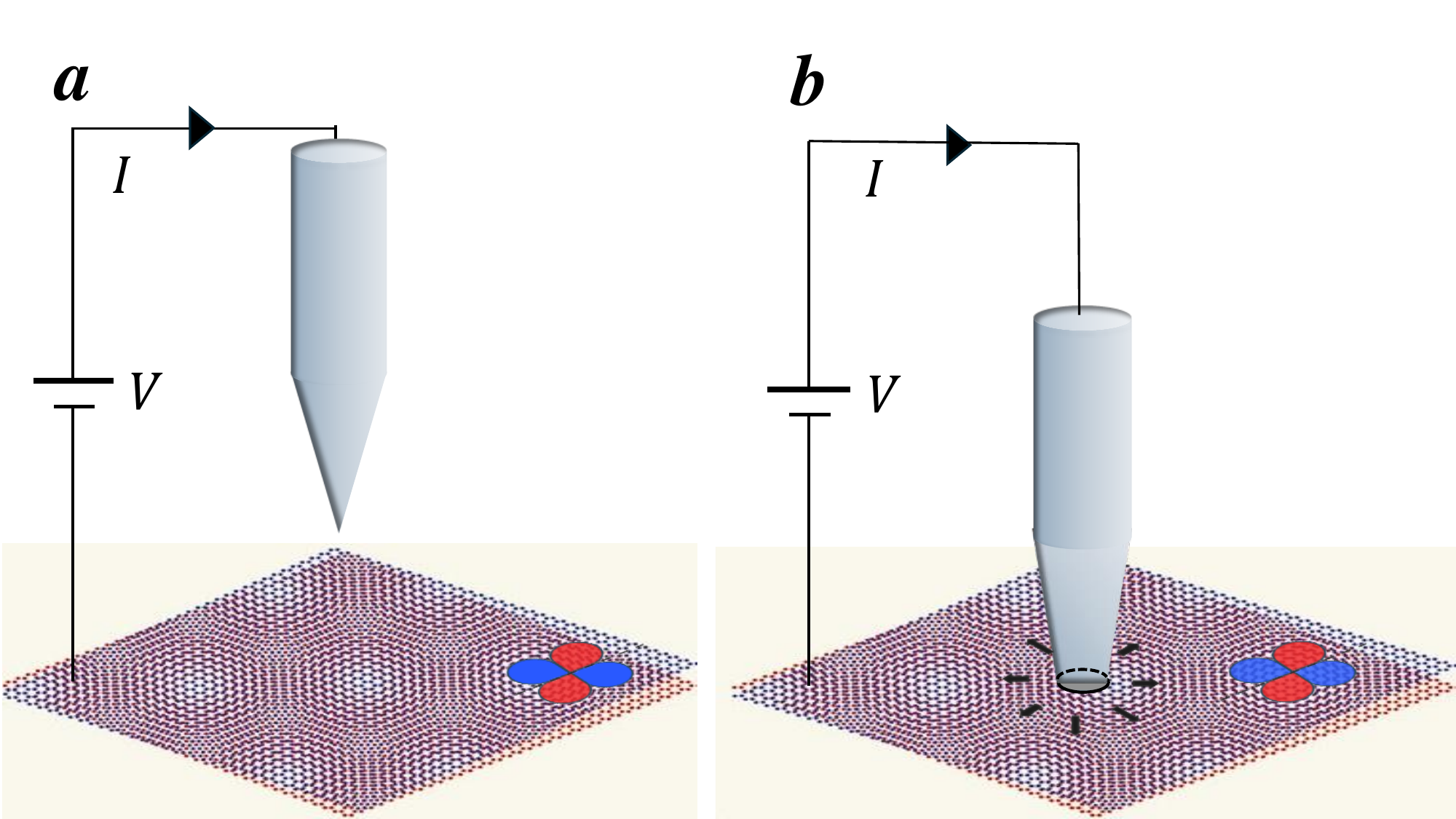}
    \caption{(a) Schematic of high-resistance tunneling experiment contrasted with (b) low-resistance Andreev experiment, which we model as a circular
			metallic disc embedded in the superconductor; see text.}
    \label{fig:cartoon}
\end{figure}

In this paper, we show that:
\\
(1) The pseudogap/SC gap interpretation of the tunneling/Andreev dichotomy is not justifiable. We show that the 
conductance in both the tunneling and Andreev regimes exhibits signatures of the {\it same} spectral features arising from the 
single-particle Green's function. This is in contradiction with the TBG and TTG experiments where the tunneling data has
no feature at the low energy scale seen in Andreev spectroscopy.
\\
(2) There is a serious problem in modeling Andreev spectroscopy in moire materials using the standard Blonder-Tinkham-Klapwijk (BTK) analysis~\cite{blonder1982} in the ballistic $Z=0$ limit. The flat bands in moire materials \cite{CastroNeto2007,bistritzer2011,kaxiras2019} imply a strong Fermi velocity $v_F$ 
mismatch between tip and sample, which effectively renormalizes~\cite{kupka1990} a transparent interface toward a tunnel barrier. 
In addition, our analysis shows that the momentum dependence of the self energy impacts the relevant $v_F$, 
but the frequency dependence does not~\cite{deutscher1994}.
\\
(3) Finally, we propose a solution that gives insight into the dichotomy between the tunneling and Andreev experiments. 
We propose that in the Andreev experiment the tip acts as a circular disc embedded in an unconventional SC. 
Our analysis of this 2D problem shows that, with strong $v_F$ mismatch, the conductance reflects the 
spectral features of the Andreev bound states (ABS) at a lower energy scale compared to the SC gap.
The ABS at the normal-SC interface arise due to interference effects in the unconventional 
SC order parameter~\cite{hu1994,kashiwaya1996,kashiwaya2000}.
\\
(4) STM in the tunneling regime probes as usual the local density of states (DOS) of the SC.
The normal-SC interface, and resulting ABS, are absent in the tunneling experiment, which then only shows the bulk SC gap.
The contrast between the tunneling and Andreev experiments is shown schematically in Fig.~\ref{fig:cartoon}.

We note that there have been several interesting theoretical papers \cite{lake2022,lewandowski2023,sukhachov2023} discussing Andreev spectroscopy in moire materials, but none of them have addressed the issues highlighted above. We comment further on comparing our results with these papers at the end.

{\bf Green's function formalism:}  
Our starting Hamiltonian is $H = H_N + H_S + H^\prime$, where 
$H_N$ describes the normal (N) metal, $H_S$ the  superconductor (S), and
$H^\prime$ couples the two. In 1D with a sharp N-S interface, or in a point contact regime
without momentum conservation, 
$H' = \gamma \sum_{kk'\sigma} c^\dagger_N(k,\sigma)c_S(k',\sigma) +\textrm{h.c.}$, 
where $c^\dagger_{j}(k,\sigma)$ creates an electron with momentum $k$, spin $\sigma$, and $j = N,S$.
(We have generalized this to interfaces in 2D but the 
simplest analysis suffices here).

To obtain the first two results highlighted in the Introduction, we
develop a Green's function approach using the Schwinger-Keldysh technique
to compute the differential conductance $g = dI/dV$ of the N-S boundary.
Both N and S are in thermal equilibrium with N kept at a potential $eV$ with respect to S.
The final answers depends only on retarded/advanced ($R/A$)
bulk N and S Green's functions, because the Keldysh $G_j^K$ is
related to $G_j^R$ via the fluctuation-dissipation theorem.

It is convenient to work in the two-component Nambu basis for both $j = N,S$,
and to define momentum-integrated Green's functions 
$\Lambda_{j}^{R}\!\left(\omega\right) = \sum_k G^{R}_{j}\!\left(k,\omega\right)\sigma_z$
and $\Lambda_{j}^{A} = \left[\Lambda_{j}^{R}\right]^\dag$, 
with a Nambu $\sigma_z$ included for later convenience.
At zero temperature we find
\beq
    g\left(V\right) = \frac{2e^2}{h} \gamma^2\,\sum_{s=\pm 1} \text{tr}\left(\Lambda^R_{\text{S}}\mathcal{M} 
    - \mathcal{M}\,\Lambda^A_{\text{S}}\right)\vert_{\omega = s e V}
\label{eq:dIdV}
\eeq
where the trace is in Nambu space and $\mathcal{M}$ is given by
$
\mathcal{M}\!=\!\left(1 - \gamma^2\Lambda^R_{N}\Lambda^R_{S}\right)^{-1}\!{\cal P}_s\!\left(\Lambda^R_{N}-\Lambda^A_{N}\right)
\!\left(1 - \gamma^2\Lambda^A_{S}\Lambda^A_{N}\right)^{-1}
$
with ${\cal P}_s = {1\over 2}\left(1+s\sigma_z\right)$. 
$\mathcal{M}$ incorporates virtual hopping processes across the N-S boundary to {\it all orders in} $\gamma$,
essential to describe the conductance down to ballistic Andreev limit.

\begin{figure}
    \centering
    \includegraphics[width=\linewidth]{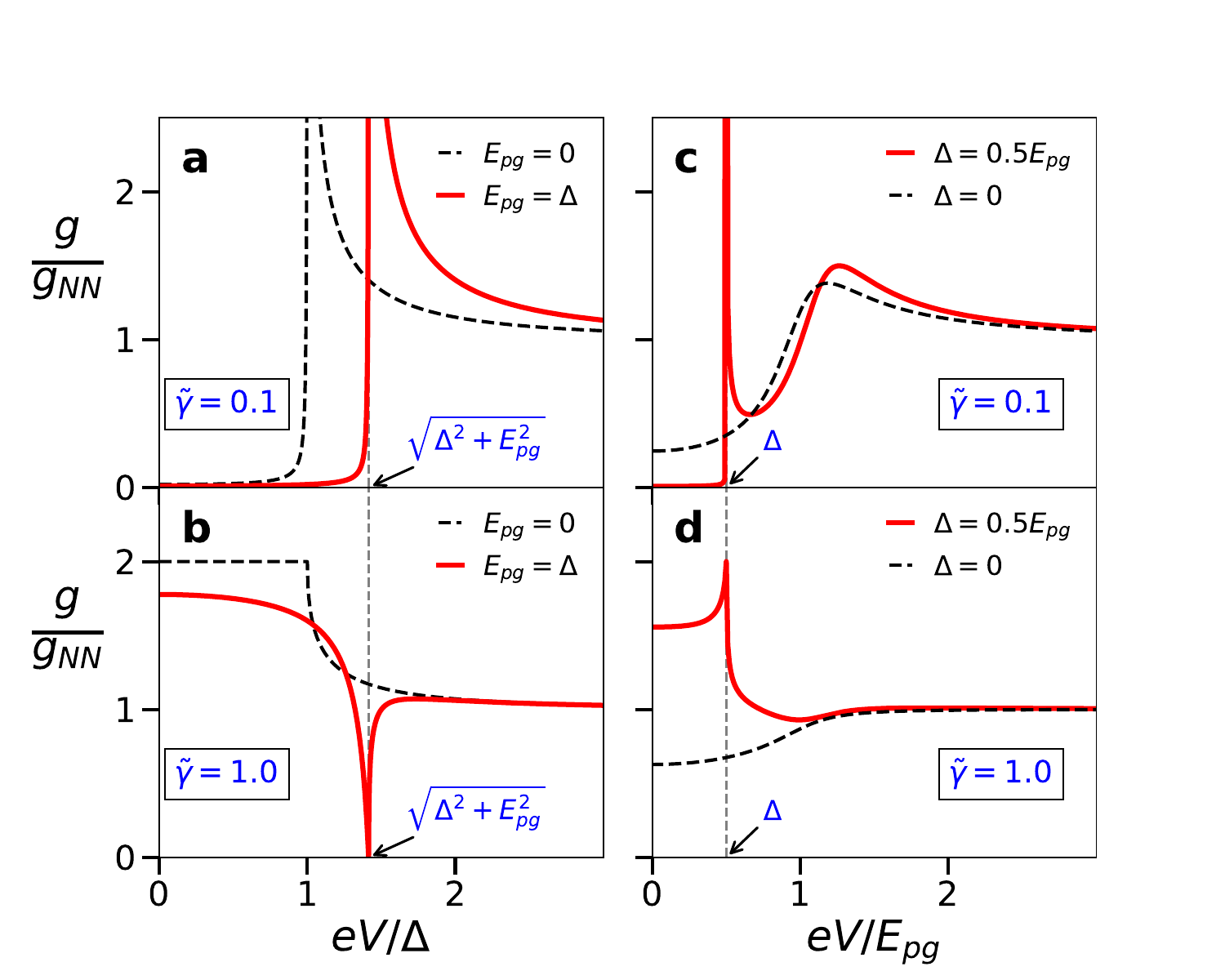}
    \caption{Differential conductance $g\left(V\right)$ normalized by its large bias value $g_{\text{NN}}$. (a) Tunneling ($\Tilde{\gamma} = 0.1$) and (b) Andreev ($\Tilde{\gamma} = 1.0$) regime results for model with a hard pseudogap $E_{\text{pg}}$ and SC gap $\Delta$ (red solid) compared with
 standard BTK results (black dashed) where $E_{\text{pg}} = 0$.   
(c) Tunneling ($\Tilde{\gamma} = 0.1$) and (d) Andreev ($\Tilde{\gamma} = 1.0$) regime results for model with a ``soft" pseudogap on scale $E_{\text{pg}}$ with 
$\Delta = 0.5E_{\text{pg}}$ (solid red) compared with pseudogap state results (black-dashed) without SC ($\Delta=0$). Note that both the 
tunneling and Andreev regimes show characteristic features at the {\it same} energy scales.}
    \label{fig:pseudogap}
\end{figure}

We note that eq.~\eqref{eq:dIdV}, whose derivation we defer to a separate publication \cite{biswas2025},  
is much more compact than equivalent expressions in the literature \cite{cuevas1996, yang2010, lake2022}.
The connection between eq.~\eqref{eq:dIdV} and the earlier literature is shown in the Supplement.
We use the BCS-Gorkov Green's function
 $G^R_S(k,\omega) = \left[\omega+i0^+ - \xi_k \sigma_z -\Delta \sigma_x\right]^{-1}$,
 with normal state dispersion $\xi_k = \epsilon_k - \mu$ and SC gap $\Delta$,
 in eq.~\eqref{eq:dIdV} to reproduce the standard BTK result \cite{blonder1982} for $g(V)$ as a function
 of $\tilde{\gamma} = \pi\gamma\rho(0)$, where $\rho(0)$ is the DOS at the chemical potential.
 This can be seen from the black dashed lines in Fig.~\ref{fig:pseudogap}(a,b),
 where panel (a) corresponds to the small $\tilde{\gamma}$ tunneling regime and (b) to the $\tilde{\gamma} = 1$ Andreev limit, as we show next.
 
To establish the relation between $\tilde{\gamma}$ and the BTK transparency $Z$ we set $\Delta=0$ 
to analyze the N-N conductance; see Supplement for details. We find
\beq
g_{\rm NN} = \left(2e^2/h\right){4\tilde{\gamma}^2}/{(1+\tilde{\gamma}^2)^2}.
\label{eq:gnn}
\eeq
Comparing this with the BTK result $\left(2e^2/h\right)/{(1+Z^2)}$, we find
$Z = \left({\tilde\gamma}^{-1} -  \tilde{\gamma}\right)/2$ for $0 < \tilde{\gamma}  \leq 1$.
The tunneling limit $Z \gg 1$ is thus obtained when $\tilde{\gamma} \ll 1$, while the Andreev regime $Z \rightarrow 0$ 
corresponds to $\tilde{\gamma} \rightarrow 1$, where the conductance reaches the ballistic Sharvin limit $2 e^2/h$. 
We analyze below the important question of how this result is affected
when the Fermi velocities, and thus the DOS, of the two metals are strongly mismatched.

{\bf Pseudogap versus SC gap:}
The first question we address is whether we can identify two different energy scales; a pseudogap  $E_{\rm pg}$ in the large $Z$ tunneling limit
 that is distinct from the coherent SC gap $\Delta$ in the $Z=0$ Andreev limit. 
Toward this end, we have used various Green's functions with two distinct energy scales and always found that 
such an interpretation is not tenable.
While the details depend upon the specific pseudogap model, the qualitative conclusions are the same. 
All regimes -- from tunneling to Andreev --
show characteristic features at the {\it same} energy scales
arising from the singularities of the Green's functions
in eq.~\eqref{eq:dIdV}.

First, consider $G_{\text{S}}(k,z) = (z^2 - \xi_k^2 - E_{\rm pg}^2 - \Delta^2)^{-1} \times \left[ z - \xi_k\sigma_z - \Delta\sigma_x\right]$, which describes a system with a 
spectral gap $E_g = \sqrt{E_{\rm pg}^2 + \Delta^2}$ distinct from the SC gap $\Delta$. The addition of 
the pseudogap and SC gap in quadrature is motivated by, e.g., refs~\cite{nozieres1999, loh2016},
which analyze the appearance of superconductivity is a normal state with ``hard" gap $E_{\rm pg}$.
We can easily incorporate anisotropies in the SC gap and in the pseudogap, but it does not affect the qualitative 
results we want to emphasize. The unconventional nature of the SC gap {\it will} play a crucial role in the last part of the paper; see below.

Using $G_{\text{S}}^{(R/A)}(k,\omega)$ obtained by $z=\omega\pm i0^+$ in eq.~\eqref{eq:dIdV} we obtain
the conductance shown in Fig.~\ref{fig:pseudogap}(a,b). 
The red curves in panel (a) show the $\tilde{\gamma}=0.1$ tunneling regime and (b) the $\tilde{\gamma} = 1$ Andreev limit result.
In both cases the conductance has a characteristic feature when the bias voltage matches the spectral gap $E_g$. (The dashed black curves in (a) and (b) are 
the standard SC state results without any pseudogap).

Another pseudogap model is one where SC develops in a normal state
DOS $\rho_{\rm pg}(\epsilon)$ with a ``soft" suppression of spectral weight on the $E_{\rm pg}$ scale;
see Supplement for details. Here we take
$\Lambda^R_{\text{S}}(\omega) = \int d\epsilon\,\rho_{\rm pg}(\epsilon)\left[z^2 - \xi^2 - \Delta^2\right]^{-1} \left[ z - \xi\sigma_z + \Delta\sigma_x\right]\sigma_z$, 
with $\xi=\epsilon- \mu$ and $z=  \omega+i0^+$.
This is inspired by and captures the basic idea in ref.~\cite{yang2010}, that focused on the more complex phenomenology of underdoped cuprates.

We plot in Fig.~\ref{fig:pseudogap}(c,d) the conductance obtained from this model in the (c) tunneling and (d) Andreev regimes. 
All curves are normalized by a constant equal to their high bias value $g_{\text{NN}}$. The black dashed curves
are the ``normal" state results with a soft pseudogap but no SC, while the red solid lines show the effect of the SC gap $\Delta$ riding on top of the pseudogap DOS.
Here we see characteristic features in both the tunneling and Andreev spectra on the scale of $\Delta$. In addition, the pseudogap shows up in tunneling as the remnant of the high energy scale at which the DOS is suppressed and in Andreev as the scale on which the conductance approaches $g_{\text{NN}}$.

Finally, we have also looked at the BCS-BEC crossover for a single band SC, 
where the spectral gap $\sqrt{\mu^2 + \Delta^2}$ is distinct from $\Delta$ once $\mu < 0$ in the BEC-like regime~\cite{Randeria1989}. 
We will report theses results separately~\cite{biswas2025}, but they lead to the same qualitative conclusion as the examples discussed above: 
the tunneling and Andreev spectra exhibit characteristic features at the {\it same} energy scales.

{\bf Velocity mismatch and self-energy effects:}
We next address how $\alpha = v_{\text{FS}}/v_{\text{FN}}$  the Fermi velocity mismatch
between the SC and the normal metal affects Andreev spectroscopy. 
Although the effect of the mismatch between {\it bare} (band structure) Fermi velocities 
has been addressed earlier~\cite{kupka1990}, its importance in the context of the flat band SCs has not been recognized.

This also raises the question of the role of interactions in renormalizing the $v_{\text{F}}$ that enters 
the interface conductance. Ref.~\cite{deutscher1994} analyzed this using a non-standard approach,
introducing non-local self energies into a wavefunction formulation. We provide more clarity on these
issues using the Green's function-based formulation of eq.~\eqref{eq:dIdV}, where self-energy effects are transparent.

Let us consider the N-N$^\prime$ interface where N$^\prime$ is the normal state of the SC.
It suffices to look at the zero bias conductance $g_{\rm NN^\prime}(0)$ as explained below.
We will see that the key parameter that determines this is the DOS ratio
$\rho_N(0)/\rho_{N^\prime}(0) = v_{\text{FS}}/v_{\text{FN}}\equiv \alpha$ for the 1D or 
point contact model we are analyzing. 
Using $\Lambda_j^{R/A}(0) = \mp i\pi \rho_\alpha(0)\,\sigma_z$ for $j = N, N^\prime$,
we find that $g_{\rm NN^\prime}(0)$ has the same form as eq.~\eqref{eq:gnn} with the crucial difference that 
now $\tilde{\gamma}^2 = \pi^2\gamma^2\rho_N(0)\rho_{N^\prime}(0)$. Effectively this means that 
we must replace $\tilde{\gamma}^2 \rightarrow \alpha \tilde{\gamma}^2$.

In the limit of small $\alpha$ this implies we are effectively in the tunneling regime. 
The upshot of this analysis is that with strong velocity mismatch one can never be in
the ballistic Andreev limit. This result is not just valid for 1D or for a point contact, 
but also holds in the 2D case as we shall demonstrate in detail below. 

Next, we turn to self energy corrections. In the Supplement, we address this question in a general manner assuming only the existence of
a Fermi surface defined by $[G^R(k_F,0)]^{-1} = 0$ without assuming sharp quasiparticles. But here, for the sake of simplicity, we focus 
on a Fermi liquid normal state. Standard arguments then lead to a single-particle Green's function 
$G^{R}\left(k,\omega \right) \approx {\mathcal{Z}} / \left( \omega+i0^+ - \mathcal{Z} \xi_k \right) + \ldots$ for
$k$ near $k_F$ and small $\omega$, where the quasiparticle weight 
${\mathcal{Z}} = 1/[1-(\partial\Sigma/\partial\omega)_{k_F,0}]$ and $\xi_k = v_F(k - k_F)$. Note that there are
{\it three} ``Fermi velocities": (i) the {\it bare} $v_F^0$ from band structure, (ii) $v_F = v_F^0 - (\partial\Sigma/\partial k)_{k_F,0}$ which includes,
e.g., Hartree-Fock corrections, and (iii) the fully renormalized $v_F^* = {\mathcal{Z}} v_F$.

The DOS at the chemical potential
$\rho(0) = -{1\over\pi}\sum_k{\rm Im}G^R(k,\omega = 0) = \sum_k {\mathcal Z}\delta( {\mathcal Z}\xi_k) = \sum_k \delta(\xi_k)$ scales like $1/v_F$.
Thus it is $v_F$ that enters the conductance calculated from eq.~\eqref{eq:dIdV}. 
$v_F$ includes the effects of k-dependent self energies, such as Hartree-Fock corrections to the bare
band structure $v_F^0$,  but {\it not} the $\omega$-dependent self energy corrections arising from
say electron-phonon or electron-electron interactions.

We note that we have focused on the zero bias conductance between two normal metals. At finite bias
one may be able to see self-energy effects beyond what we discuss here \cite{lee2015}. It suffices, however, for our purposes to look at 
$g_{\rm NN^\prime}(0)$ to determine the scale of the conductance, and from this to conclude that one cannot remain in the ballistic limit 
for strong velocity mismatch. 

{\bf Metallic disc in a $d$-wave SC:} Having concluded (1)  that the two different energy scales seen in tunneling and Andreev spectroscopies cannot be
attributed to the pseudogap and the SC gap, and (2) that a transparent barrier is renormalized towards the tunneling regime when the 
the Fermi velocities are strongly mismatched, we must now ask how we can understand the observed Andreev spectra in TBG and TTG. 

We model the STM tip that has crashed into the sample as a circular normal metal disc of radius $R$
embedded in an unconventional 2D SC ($r > R$). As a concrete example, we consider a $d$-wave SC with $\Delta(\phi) =  \Delta_0 \cos(2\phi)$ 
which also supports nodal quasiparticles as suggested by the experiments noted in the Introduction. The sign change in the order parameter 
leads to the appearance of Andreev bound states (ABS)~ \cite{hu1994,kashiwaya1996,kashiwaya2000} at the N-S interface which will play a crucial role in our analysis.

In the spirit of looking at the simplest model that captures the qualitative physics, we describe
the N and S electronic dispersion with parabolic bands with masses $m_{\text{N}}$ and $m_{\text{S}}$, and Fermi wave-vectors $k_{\text{FN}}$ and $k_{\text{FS}}$,
We have found that our results depend weakly on $\zeta = k_{\text{FS}}/k_{\text{FN}}$ but, as emphasized below,
exhibit a strong dependence on $\alpha = v_{\text{FS}}/v_{\text{FN}} = \zeta (m_{\text{N}}/m_{\text{S}})$.

We found it illuminating to analyze this 2D problem using the Bogoliubov-de Gennes (BdG) formalism. 
We consider a source at the origin that emits an outgoing wave with energy $E$ incident upon the N-S interface with a bare $Z\!=\!0$. 
We work with partial waves labeled by angular momentum $l_0$, determine the scattering amplitudes for each $l_0$, and sum over $l_0$ to
obtain the conductance. 

The partial wave for given $l_0$ will be reflected and transmitted across the interface with $l$-values that can differ from  
$l_0$ by an even integer for a $d$-wave SC. 
The normal metal wavefunction for $0< r < R$ is
\begin{widetext}
\beq
\Psi(r,\theta) = i^{l_0}e^{il_0\theta}\,H^{(1)}_{l_0}(k_{\text{FN}}  r) \chi^{p}_{N}
        + \sum_{l=l_0+2n} i^{l}e^{il\theta} \left(b_l H^{(2)}_{l}(k_{\text{FN}}  r) \chi^{p}_{N}        
         + a_l H^{(1)}_{l}(k_{\text{FN}}  r)  \chi^{h}_{N}\right)
\label{eq:psi-N}
\eeq
 \end{widetext} 
The  dependence of the amplitudes $a_l$ and $b_l$ on $E$ and $l_0$ is suppressed for simplicity.
Here $H_l^{\left(i\right)}$ are Hankel functions and $\chi^{p}_{N} = \left(1,0\right)^T$ and $\chi^{h}_{N} = \left(0,1\right)^T$ are particle and hole spinors.
The SC wavefunction for $r > R$ is
\begin{widetext}
 \beq
 \Psi(r,\theta) = \int_0^{2\pi} \frac{d\phi}{2\pi} \left( c(\phi) \sum_{l = l_0 + 2n} i^l e^{il(\theta-\phi)} H^{(1)}_l(k_{\text{FS}}r) \chi^{p}_{S}
+ d(\phi) \sum_{l = l_0 + 2n} i^l e^{il(\theta-\phi)} H^{(2)}_l(k_{\text{FS}}r) \chi^{h}_{S} \right).
\label{eq:psi-S}
\eeq
 \end{widetext} 
Here $\chi^{p}_{S} = \left( u(\phi), v(\phi)\right)^T$ is the particle-like solution of the BdG equation
$\left([{E^2-\left|\Delta(\phi)\right|^2}]^{1/2}\sigma_z +  \Delta(\phi)\sigma_x\right) \chi^{p}_{S} = E \chi^{p}_{S}$, and
$\chi^{h}_{S} = \left( v(\phi), u(\phi) \right)^T$ is the corresponding hole-like solution, with
$ \Delta(\phi) =  \Delta_0 \cos(2\phi)$.

It is convenient to work in the $l$-basis by rewriting eq.~\eqref{eq:psi-S} using convolutions like
$\int_{0}^{2\pi}\!{d\phi}\,e^{-il\phi} u(\phi)c(\phi)/{2\pi}  = \sum_{l'} u_{l-l'} c_{l'}$ in all the integrals. 
Note that $u_{l-l'}$ and $v_{l-l'} \neq 0$ only when $l-l'$ is an even integer. 

\begin{figure}
    \centering
    \includegraphics[width=\linewidth]{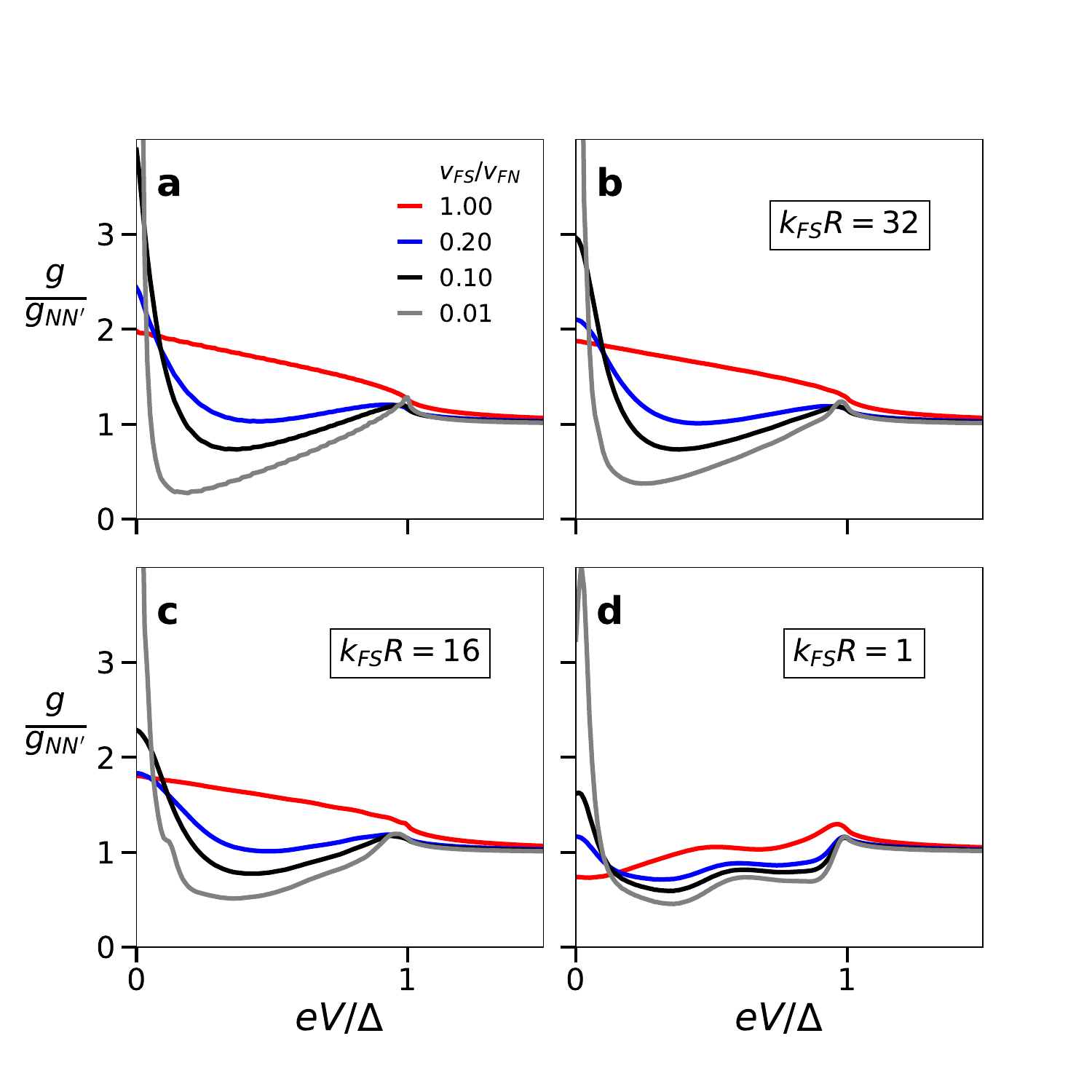}
    \caption{ Normalized differential conductance of an interface between a metal and $d$-wave SC in 2D: (a) planar junction averaged over all angular orientations of the order parameter and (b,c,d) circular junction arising from a circular metallic disc of radius $R$ embedded in the SC.  (b) $k_{\text{FS}}R =32$, (c) $k_{\text{FS}}R =16$, (d) $k_{\text{FS}}R =1$ for $\alpha = v_{\text{FS}}/v_{\text{FN}}$ equal to 1.0 (red), 0.2 (blue), 0.1(black) and 0.01(gray). $k_{\text{FS}}/k_{\text{FN}}$ is set to 0.1. 
Note that $k_{\text{FS}}R$ is a proxy for the number of Landauer channels, and as the velocity mismatch varies one goes from a ballistic Andreev limit ($\alpha = 1$)
to a tunneling regime ($\alpha \ll 1$).}
    \label{fig:circular-junc}
\end{figure}

For given $l_0$ and $E$, the boundary conditions $\Psi(R^-,\theta) = \Psi(R^+,\theta)$ and 
$(1/m_{\text{N}}) \partial\Psi(R^-,\theta)/\partial r\!=\!(1/m_{\text{S}}) \partial\Psi(R^+,\theta)/\partial r$ at the N-S interface
lead to 
\begin{eqnarray}
\delta_{ll_0}\begin{bmatrix}
1\\0\\1\\0
    \end{bmatrix} = \begin{bmatrix}
        0&B&C^u&D^v\\
        -\mathbb{I}&0&C^v&D^u\\
        0&{\bar B}&\alpha{\bar C}^u&\alpha{\bar D}^v \\
        -\mathbb{I}&0&\alpha{\bar C}^v&\alpha {\bar D}^u
    \end{bmatrix}_{ll'}
    \begin{bmatrix}
        a\\ b\\c\\d
    \end{bmatrix}_{l'}
\label{eq:matrix}
\end{eqnarray}
with an implied sum over $l'$.
The various matrix entries are given by
$B_{ll'} = - \delta_{ll'} {H^{(2)}_l(k_{\text{FN}}R)}/{H^{(1)}_l(k_{\text{FN}}R)}$, 
$C^u_{ll'} = E_l(R) u_{l-l'}$, $C^v_{ll'} = E_l(R) v_{l-l'}$, 
$D^u_{ll'} = F_l(R) u_{l-l'}$, and $D^v_{ll'} = F_l(R) v_{l-l'}$
where we have defined
$E_l(R) = {H^{(1)}_l(k_{\text{FS}}R)}/{H^{(1)}_l(k_{\text{FN}}R)}$ and
$F_l(R) = {H^{(2)}_l(k_{\text{FS}}R)}/{H^{(1)}_l(k_{\text{FN}}R)}$.
The terms with ``bars" are obtained from their ``unbarred" counterparts by replacing each Hankel function with its
derivative ${H_l^{(i)}}^\prime(z) = dH_l^{(i)}(z)/dz$, so that, e.g., ${\bar C}^u$  and ${\bar C}^v$ involve 
${\bar E}_l(R) = {{H_l^{(1)}}^\prime(k_{\text{FS}}R)}/{{H_l^{(1)}}^\prime(k_{\text{FN}}R)}$.

We need to solve eq.~\eqref{eq:matrix} for the amplitudes $a_l$ and $b_l$ corresponding to 
each incident $l_0$ as function of the energy $E=eV$. We then obtain the
conductance
\beq
        g(V) = \frac{2e^2}{h}\sum_{l_0}\left[1+\sum_l \left(|a_l(l_0,eV)|^2 - |b_l(l_0,eV)|^2 \right)\right].
\label{eq:g2d}
\eeq
The infinite sum $g= \sum_{l_0}g_{l_0}$ has a natural cutoff as
$g_{l_0}$ is strongly suppressed beyond $\l_0^{\rm max} \sim {\cal O}(k_F R)$, where
$k_F = \min\left(k_{\text{FS}},k_{\text{FN}}\right)$. We note that $(2k_{\text{F}}R+1)$ is the 
the number of channels (in the sense of Landauer theory~\cite{buttiker1985}) in an interface of circumference $2\pi R$.
We see this clearly from the computed $g\left(V\right)$ for $eV\gg\Delta$, which saturates to 
$g_{\text{NN}'} \approx \left(2k_{\text{F}}R+1\right) \left({2e^2}/{h}\right)\left[{4\alpha}/{\left(1+\alpha\right)^2}\right]$.
Note that the velocity mismatch $\alpha$ decreases the conductance of each channel below the Sharvin limit,
just as we had deduced in the 1D or point contact limit above.


In Fig.~\ref{fig:circular-junc} we show the conductance obtained from eq.~\eqref{eq:g2d}.
We begin with the results for $k_{\text{F}}R\gg1$ shown in Fig.~\ref{fig:circular-junc}(b), where $k_{\text{F}}R$ should be thought of as a proxy for number of channels.
In the $\alpha =1$ ballistic Andreev regime (red curve in Fig.~\ref{fig:circular-junc}(b)), 
the normalized conductance gradually decreases from $2$ to $1$ with increasing bias on the scale of 
$\Delta$. This is due to the angular variation of $|\Delta\left(\phi\right)|$, in contrast to the flat value of $2$ in \textit{s}-wave SC.

The sign change in $\Delta\left(\phi\right)$ gives rise to interference effects that produce Andreev bound states (ABS) at the
interface~\cite{hu1994,kashiwaya2000}. The ABS strongly modify the low-energy local DOS at $r=R$, which leads to the 
zero bias conductance peaks (ZBCP) seen in the $\alpha = 0.1$ (black) and $\alpha = 0.01$ (gray) results in Fig.~\ref{fig:circular-junc}(b). 
We also clearly see that as the velocity mismatch grows the conductance evolves from the Andreev regime ($\alpha=1$) towards
tuneling (($\alpha \ll1$).

Tunneling signatures of ABS at planar junctions have been studied both experimentally \cite{covington1996,covington1997}.
and theoretically~\cite{hu1994,fogelstrom1997,kashiwaya1996,kashiwaya2000} in the cuprates. These were analyzed in detail~\cite{kashiwaya1996,kashiwaya2000} 
as a function of the angle $\beta$ between the interface normal and the $d$-wave order parameter.
The planar junction conductance averaged over all orientations $0\leq \beta < 2\pi$ (see Supplement for details)
shown in Fig.~\ref{fig:circular-junc}(a) is in semiquantitative agreement with the results of the 
$k_{\text{F}}R\gg1$ circular interface shown in Fig.~\ref{fig:circular-junc}(b).

We turn to smaller values of $k_{\text{F}}R$ in Fig.~\ref{fig:circular-junc}(c,d), where we cannot approximate the results using known planar junction results. 
For equal velocities $\alpha=1$, we see from panels (b,c,d) that the normalized $g(0)$ decreases with decreasing $k_{\text{F}}R$,
and is smaller than unity for $k_{\text{F}}R = 1$. This trend is consistent with the vanishing Andreev signal 
for a point contact $k_{\text{F}}R\ll 1$ in a $d$-wave SC \cite{sukhachov2023}. 

In the regime of strong velocity mismatch relevant to flat band SCs, however, the signatures of the ABS persist 
down to the $k_{\text{F}}R \sim 1$ limit. 
The ABS give rise to the low energy scale observed in the Andreev experiments, which is much smaller
than the SC gap seen in tunneling spectroscopy.

We conclude by estimating $\alpha=v_{\text{FS}}/v_{\text{FN}}$ for moire SCs.
Using $v_{\text{FS}} \sim\!1-4 ~\times 10^4\,\text{ms}^{-1}$ for magic angle TBG~\cite{Kaxiras2024},
and $v_{\text{FN}} \sim\!10^5 - 10^6\, \text{ms}^{-1}$ for the metal (e.g., tungsten~\cite{Kollar1978}),
we obtain a rough estimate $\alpha \sim 0.01 - 0.25$. 

{\bf Comparison with other work:} Andreev spectroscopy in TBG and TTG was discussed recently in the context of constraints on the 
order parameter symmetry~\cite{lake2022}, and of the BCS-BEC crossover~\cite{lewandowski2023}. 
The 2D problem was, however, approximated as a sum of independent 1D BTK channels for each $\phi$,
which does not include the interference effects leading to the ABS that are central to our results.

Refs.~\cite{lake2022,sukhachov2023} emphasized that the Andreev signal is completely suppressed for a point contact in a \textit{d}-wave SC. 
This is consistent with our analysis. In our Green's function formalism, this is seen from the vanishing of the off diagonal terms in the 
$k$-integrated SC Green's function. In addition, our circular disc results at $\alpha = 1$ also show that $g\left(0\right)$ decreases as we decrease $k_{\text{F}}R$.

The DOS arising from ABS localized near defects or boundaries in \textit{d}-wave SCs has long been understood. 
In addition to the papers already cited, we note two limiting cases: the semiclassical analysis  ($k_{\text{F}}R \gg 1$) of a
strong extended defect~\cite{adagideli1999} and the $T$-matrix analysis of strong point impurity~\cite{balatskty2006} ($k_{\text{F}}R \ll 1$).
The analysis presented here goes beyond these works by analyzing the signatures of ABS in the 
differential conductance of a circular N-S interface for various $k_{\text{F}}R$ as a function of the velocity mismatch between tip and sample, which 
is crucial for moire graphene SCs.

{\bf Conclusions and open questions:}  Our main conclusions (1) through (4) are already highlighted in the Introduction. 
Our results rule out earlier interpretations in terms of pseudogap/SC gap, clarify the role of velocity mismatch between tip and sample,
and highlight the role of Andreev bound states created at the N-S interface in an unconventional SC. 
The single-band $d$-wave SC used in our analysis is only a simple example of an unconventional order parameter.
Elucidating the symmetry of the SC order parameter in TBG and TTG remains an important open question with
a sharp answer independent of many microscopic details. Andreev spectroscopy is one of the few phase sensitive experiments that 
have been used to probe TBG and TTG, and as such it would be very important to model this for a multi-band 
SC with an unconventional order parameter.

\medskip
\bigskip
{\bf Acknowledgements} M.R. acknowledges a useful conversation with Ali Yazdani. SS and RS acknowledge funding from Department of Atomic Energy, Govt. of India.

 \bibliography{Andreev_v5} 
 	
\end{document}